**Physics-Aware Machine-Learning-Driven Inverse Design of Broadband Ultra-Open Acoustic Metamaterials**

*Zhiwei Yang†, Mengyu Li†, Xiaohang Xie, Ao Chen, Thomas G. Bifano, Xin Zhang**

Zhiwei Yang, Mengyu Li, Xiaohang Xie, Ao Chen, Prof. Thomas G. Bifano, Prof. Xin Zhang
Department of Mechanical Engineering, Boston University, Boston, Massachusetts 02215, USA
Email: xinz@bu.edu

Zhiwei Yang, Mengyu Li, Xiaohang Xie, Ao Chen, Prof. Thomas G. Bifano, Prof. Xin Zhang
Photonics Center, Boston University, Boston, Massachusetts 02215, USA

Prof. Thomas G. Bifano, Prof. Xin Zhang
Division of Materials Science and Engineering, Boston University, Boston, MA 02215, USA

Prof. Thomas G. Bifano, Prof. Xin Zhang
Department of Electrical and Computer Engineering, Boston University, Boston, MA 02215, USA

Prof. Thomas G. Bifano, Prof. Xin Zhang
Department of Biomedical Engineering, Boston University, Boston, MA 02215, USA

Abstract: Ventilated acoustic silencers combing sound attenuation with high ventilation are pivotal for advanced noise control. However, balancing attenuation, bandwidth, openness, and thickness remains a high-dimensional challenge. Here, we report a physics-aware machine-learning-driven inverse design framework for ultra-open acoustic silencers (UAS). By leveraging Green's function-based parameterization, we physically decouple the design space into spectral and radial parameters, ensuring physical interpretability while reducing complexity. We introduce a two-stage forward prediction architecture that captures broadband envelopes and sharp resonant features via a coarse-to-fine strategy. Coupled with a population-based, hybrid-objective parallel (PHP) inverse strategy, our framework enables rapid exploration of non-convex landscapes, identifying hundreds of optimized candidates within seconds. Crucially, this framework uncovers hidden linear design rules that govern high-performance monolithic designs, acting as geometric proxies for optimal impedance-matching. We experimentally validate a family of prototypes: UAS-2 demonstrates the monolithic limit with high ventilation ratio, while UAS-3 demonstrates versatility in multi-mode interactions.

To circumvent the trade-off ceiling of single-unit resonators, a parallel-composite architecture (UAS-4) is introduced to enhance performance through spatial interference distribution. Results confirm a broadband bandwidth exceeding 830 Hz achieved with an ultra-thin profile (0.1–0.2$\lambda$) and 80% ventilation. This work establishes a data-driven paradigm for discovering design principles in functional metamaterials.

Zhiwei Yang and Mengyu Li contributed equally to this work.

## 1. Introduction

Noise control is a pervasive challenge in modern society, especially in scenarios demanding both effective airborne sound attenuation and high airflow permeability.[1-4] Conventional sound silencers, including Helmholtz resonators, membrane absorbers, Fabry-Pérot (FP) resonators, and Fano resonators, typically achieve robust acoustic suppression at the cost of obstructing fluid transport, creating an inherent conflict between noise reduction and ventilation.[5–15] To reconcile these competing requirements, ventilated silencers, such as meta-muffler, meta-cage, Meta-barrier, phase-gradient metasurfaces, Bragg-scattering structures, Fano resonators, and hybrid-type resonances, have been extensively explored.[16–33] Among these, bilayer Fano-resonant systems stand out by leveraging destructive interference between discrete resonant pathways and a continuous background to produce deep transmission dips while maintaining open channels. However, the efficacy of such local resonances is fundamentally constrained by causality-governed trade-offs: broadening the attenuation bandwidth or deepening the suppression level typically necessitates increased structural complexity, which inevitably compromises structural openness and subwavelength compactness.[34–36]

Navigating these multidimensional trade-offs requires exploring a vast, non-convex design space where geometric and physical parameters are intricately coupled. Traditional design paradigms, predominantly rooted in empirical rules or intuition-guided trial-and-error, are computationally prohibitive and often fail to uncover global optima within such high-dimensional design space.[37–39] While machine learning (ML) has recently accelerated metamaterial discovery via high-speed surrogate modeling, two critical bottlenecks persist in its application to resonant acoustics.[40–49] First, standard end-to-end neural networks often struggle to resolve broad spectral trends and sharp, high-Q resonant features simultaneously, frequently resulting in spectral blurring or misalignment.[50–53] Second, the inverse design landscape is hindered by the fundamental "one-to-many" mapping problem, where distinct topologies yield degenerate physical responses. Conventional optimizers often converge to localized solutions, overlooking diverse candidate regions that might better satisfy auxiliary engineering constraints such as ventilation ratio or fabrication feasibility.[54–57]

To address these challenges, we report a physics-aware, machine-learning-assisted inverse design framework for the deterministic discovery of ultra-open acoustic silencers (UAS) based on bilayer Fano resonances (**Figure 1**). By integrating a Green's function with effective medium theory, we physically decouple the design space into spectral and radial parameters, effectively encoding the underlying scattering physics into the learning process while reducing design space complexity. We implement a two-stage "coarse-to-fine" surrogate architecture that progressively refines spectral accuracy, capturing both broadband envelopes and sharp Fano features with high precision. Based on this surrogate, a population-based hybrid-objective parallel (PHP) inverse prediction with stage-wise survivor selection is employed to comprehensively navigate the high-dimensional landscape. This "many-to-many" strategy identifies hundreds of candidate structures satisfying the target transmission spectra, requiring only seconds per generated design to allow for the precise tailoring of attenuation profiles. Crucially, our framework uncovers hidden linear design rules that serve as geometric proxies for optimal impedance-matching, providing a deterministic "ruler" for scaling ultra-open structures. Based on these rules, we identify four distinct classes of UAS (UAS-1 to UAS-4) that substantially surpass the performance limits of existing ventilated metamaterials. We show that monolithic designs (UAS-2 and UAS-3) push the boundaries of single-unit efficiency, the parallel-composite architecture (UAS-4) effectively circumvents the monolithic trade-offs through the spatial distribution of interference sites. Experimental validation confirms an exceptional broadband silencing bandwidth exceeding 830 Hz (within the 1-2 kHz range) achieved with an ultra-thin profile (0.1-0.2 $\lambda$) and ventilation ratios up to 80%. These results establish a robust, data-driven paradigm for the discovery of design principles in functional acoustic metamaterials.

## 2. Inverse Design Framework

### 2.1. Physics-informed Descriptors and Dataset Construction

A critical step of the proposed framework is the construction of physics-informed descriptors that encode both geometric configurations and acoustic response of the ultra-open silencers (UAS). Building upon the bilayer scattering system and effective medium theory (EMT), the UAS is conceptualized as a dual-path acoustic interferometric system, as illustrated in Figure 1b. The structure geometry is parameterized by the waveguide radius $R$, the inner ($j = 1$) and outer ($j = 2$) radii $r_{i,j}$ of each region ($i = 1, 2$), and the axial thickness $t$. Each region is further characterized by its effective acoustic refractive index $n_i$ and acoustic impedance $Z_i = \rho_i c_i$. Region 1 forms a straight air channel supporting a transmission continuum, while Region 2 serves as a phase-delayed propagation pathway. The interference between these discrete and continuous pathways governs the emergence of Fano resonances.

In this framework, $n_2$ and $t$ are categorized as spectral parameters, as they predominantly dictate the accumulated acoustic phase $\phi \sim n_2 t$ . Conversely, $r_{i,j}$ are treated as radial parameters primarily governing modal coupling strengths, consistent with the Green's function-based scattering model (see Discussion for formal derivation). While these two groups of parameters exhibit weak physical coupling within the underlying scattering physics, their functional roles in modulating the transmission profiles are distinct and dominant. This functional decoupling effectively reduces the design space complexity while preserving physical interpretability, providing a robust input for the machine-learning framework to efficiently navigate the UAS design space and prediction of transmission spectra.

To construct the design-response dataset, UAS configurations are randomly sampled within the geometrically feasible domain of this descriptor space (Figure 1c, see Supplementary Information for details). For each configuration, the boundary acoustic fields are evaluated using Green's function of the bilayer system. Let $P_{i0}$ ($P_{it}$) and $U_{i0}$ ($U_{it}$) denote the acoustic pressure and volume velocity at the inlet (outlet) of region $i$, respectively. The power transmission coefficient $T$ of the bilayer system is defined as:

$$T = \left| \frac{\rho_0 c_0 (U_{1t} + U_{2t})}{\pi R^2} \right|^2$$

where $\rho_0$ and $c_0$ are the density and sound speed of the ambient medium (see Supplementary Information for details). The transmission spectrum of each configuration is computed over a prescribed frequency grid, forming a structured dataset of 200,000 design-response pairs. This scale ensures that the neural network can resolve the sharp gradients in the design landscape associated with high-Q Fano resonances. This extensive dataset facilitates the training of machine-learning architecture (Figure 1d, e), which comprises a two-stage forward-prediction network and a population-based hybrid-objective parallel inverse prediction strategy.

**2.2. Forward Prediction with a Two-stage Training Architecture**

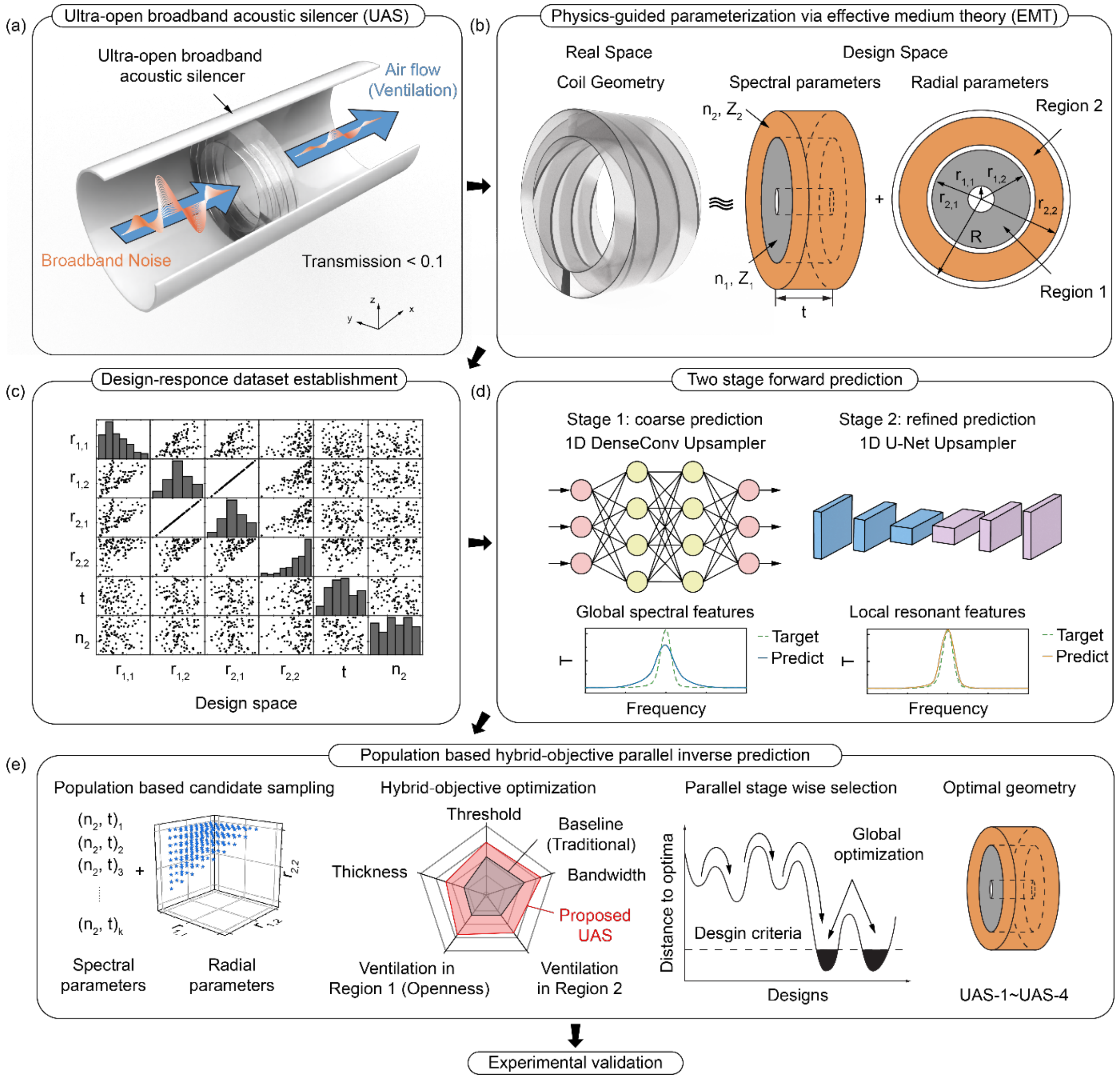

**Figure 1. Machine learning (ML)-based inverse design framework for ultra-open acoustic silencers (UAS).** (a) Schematic of the UAS, showing effective broadband noise suppression with simultaneous high airflow permeability (ventilation). (b) Physics-informed parameterization of the bilayer scattering system via effective medium theory (EMT), where the configuration is decoupled into spectral parameters ($n_2$, $t$) and radial parameters ($r_{1,1}$, $r_{1,2}$, $r_{2,1}$, $r_{2,2}$). (c) Two-stage forward prediction architecture consisting of a 1D DenseConv-based coarse prediction and a 1D U-Net-based refined prediction to resolve sharp resonant features. (e) Population-based hybrid-objective parallel inverse prediction strategy, encompassing candidate sampling within the design space, multi-objective optimization (bandwidth, ventilation, and compactness), and stage-wise survivor selection for global optimization of UAS configurations.

To establish an accurate mapping from the six-dimensional design descriptor

$$x = \{r_{1,1}, r_{1,2}, r_{2,1}, r_{2,2}, t, n_2\} \in \mathbb{R}^6$$

to its frequency-domain transmission spectrum, we develop a two-stage one-dimensional (1D) hierarchical neural network **(Figure 2a)**. The targeted transmission spectra exhibit distinct multi-scale characteristics, combining smooth, broadband global trends with sharp, high-Q localized resonances stemming from Fano interference. This intrinsic physical complexity motivates a tailored coarse-to-fine learning strategy. In conventional single-stage end-to-end training, the optimization process invariably prioritizes the dominant smooth trends, often leading to the unintended smoothing of narrow resonance peaks or positional misalignment of critical spectral features.

The proposed network architecture consists of a Dense Convolution Upsampler (DenseConv) for Stage 1, serving as a coarse predictor, followed by a U-Net Upsampler refinement module (U-Net) for Stage 2 (Figure 2a). In Stage 1, the low-dimensional design descriptor is first projected and reshaped into an initial feature representation, which is then progressively upsampled through four transposed-convolution blocks to produce a high-resolution latent embedding of size $64 \times 256$. Each upsampling block integrates dense convolutional layers coupled with a channel-wise attention module to adaptively enhance feature representation. The resulting latent embedding is linearly projected to generate an initial prediction of the transmission spectrum, successfully capturing the global spectral features. Upon convergence of Stage 1, this latent embedding is passed to Stage 2 for high-fidelity refinement. Stage 2 employs a 1D U-Net architecture, leveraging its intrinsic skip connections to operate on the Stage 1 latent embedding, thereby efficiently recovering localized, high-frequency spectral features while preserving global consistency. The refined embedding is subsequently flattened and projected through a multilayer perceptron (MLP) head to output the final, high-fidelity predicted transmission spectrum (see Supplementary Information for details). Training is executed via a sequential two-stage curriculum strategy to ensure stable convergence across scales. Stage 1 is trained first for $N_1$ epochs while Stage 2 remains frozen. Subsequently, Stage 1 is frozen, and Stage 2 is trained for $N_2$ epochs to refine the latent representation. Both stages utilize a mean squared error (MSE) loss function and the AdamW optimizer, with $N_1 = N_2 = 1000$. The established dataset of 200,000 configurations is partitioned into 80% training and 20% testing sets. All design variables are normalized prior to training to ensure a conditioned optimization landscape.

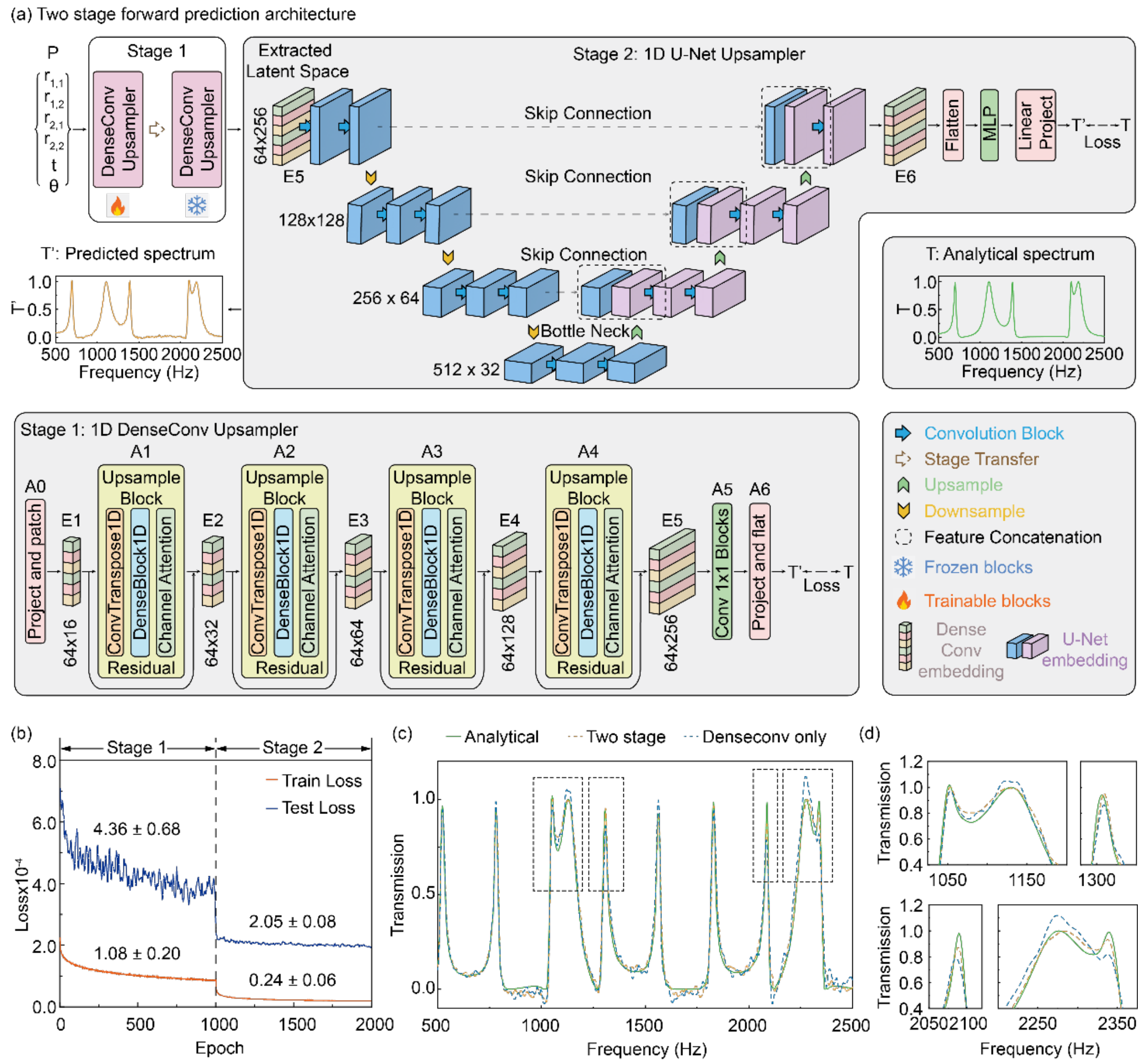

**Figure 2. Two-stage hierarchical forward prediction architecture and performance.** (a) Detailed architecture of the forward prediction network, featuring a Dense Convolution Upsampler (DenseConv, Stage 1) for global feature learning and a 1D U-Net Upsampler (U-Net, Stage 2) for high-frequency spectral refinement. Flame and snowflake icons denote trainable and frozen blocks, respectively, illustrating the sequential curriculum training strategy. The low-dimensional design vector is progressively upsampled and processed through channel-wise attention and skip connections to output the transmission spectrum $T'$. (b) Training and test loss evolution across the two stages. The vertical dashed line marks the transition where the DenseConv is frozen and the U-Net is activated, leading to a substantial reduction in both loss magnitude and variance. (c) Full-spectrum comparison for a representative test configuration, illustrating the superior fidelity of the two-stage model (orange dashed line) relative to the DenseConv-only baseline (blue dotted line) in matching the analytical spectra (green solid line). (d) Zoomed-in views of the four spectral regions highlighted by dashed boxes in (c), demonstrating the framework's capability to precisely reconstruct sharp Fano resonances and steep spectral slopes that are typically smoothed in single-stage architectures.

The training performance of the hierarchical network is summarized in Figure 2b. During Stage 1, both the training loss ($(1.08 \pm 0.20) \times 10^{-4}$) and test loss ($(4.36 \pm 0.68) \times 10^{-4}$) decrease steadily yet exhibit noticeable fluctuations, underscoring the inherent instability of DenseConv alone in capturing sharp spectral features. Upon activating Stage 2 refinement, both losses plummet substantially to $(0.24 \pm 0.06) \times 10^{-4}$ (training) and $(2.05 \pm 0.08) \times 10^{-4}$ (test), accompanied by a markedly reduced variance. Quantitatively, the two-stage strategy reduces the standard deviation of the training loss by 88% and that of the test loss by 70% compared to the DenseConv-only baseline. These results substantiate that staged refinement drastically improves convergence stability while simultaneously enhancing predictive accuracy for multi-scale resonance spectrum.

Representative predictions further validate these performance gains (Figure 2c, d). Figure 2c compares the transmission response synthesized by the analytical bilayer model, the hierarchical two-stage model, and the DenseConv-only baseline for a representative test sample. While both models replicate the overall spectral envelope, pronounced discrepancies arise in regions characterized by sharp and closely spaced resonances. The DenseConv-only baseline captures the broad spectral structure but suffers from smoothed peaks, misalignment of resonance positions, and occasional spurious oscillations. Conversely, the hierarchical two-stage model precisely reconstructs narrow resonances and steep spectral slopes, exhibiting exceptional agreement with the analytical spectrum across both smooth and rapidly varying regimes. The zoomed-in views in Figure 2d further highlight the superior peak-level spectral fidelity achieved by the two-stage architecture. These results confirm that the DenseConv module primarily learns the coarse spectral space representing the global response structure, while the U-Net refinement stage faithfully restores fine-scale spectral features from the latent embedding. The resulting coarse-to-fine hierarchy enables robust reconstruction of sharp resonance transitions without sacrificing global consistency.

### 2.3. Population-based Hybrid-objective Parallel (PHP) Inverse Prediction Framework

Leveraging the trained hierarchical two-stage forward surrogate model, the inverse design problem is formulated as predicting optimal physical parameters that satisfy a prescribed target transmission spectrum. Given a target response $T$, the objective is to determine a six-dimensional design vector $x \in \mathbb{R}^6$ such that the surrogate prediction $\hat{T}(x) = g_{\phi}(x)$ satisfies band-limited performance constraints.

To enable consistent mechanisms-driven comparison across diverse geometric configurations and eliminate trivial frequency translation during optimization, the spectral parameters $\{n_2, t\}$ are fixed to predefined nominal values $\{n_{20}, t_0\}$. This strategic treatment allows the inverse prediction to focus exclusively on optimizing the radial coupling mechanisms for potent attenuation while maintaining a unified normalized frequency coordinate. Accordingly, the physical frequency axis $f$ is transformed into a normalized frequency $f'$ via:

$$f' = \frac{n_{20} t_0}{c_0} f$$

The inverse objective is exclusively evaluated only within a prescribed normalized frequency range $f' \in [f'_{low}, f'_{high}]$. The primary design objective is to ensure that the predicted transmission spectrum remains below a predefined suppression threshold $\tau$ within the target band. To achieve this, a band-limited overshoot loss is defined as:

$$\mathcal{L}_{over}(\boldsymbol{x}) = \sum_{k \in \mathcal{B}} \max\left(\widehat{T_k}(\boldsymbol{x}) - \tau, 0\right) \Delta f'_k,$$

where $\mathcal{B}$ denotes the index set of discrete frequency samples within the optimization band.

This loss specifically penalizes spectral regions that violate the attenuation threshold, avoiding unnecessary penalties in already-suppressed regions. To prevent convergence toward overly conservative, degenerate solutions where transmission is excessively suppressed far below the threshold, potentially sacrificing ventilation openness or effective bandwidth, a complementary margin term regularization is introduced:

$$\mathcal{A}_{margin}(\boldsymbol{x}) = \sum_{k \in \mathcal{B}} \max\left(\tau - \widehat{T_k}(\boldsymbol{x}), 0\right) \Delta f'_k$$

The final optimization objective is defined as a hybrid composition:

$$\mathcal{J}(\boldsymbol{x}) = \lambda_{over} \mathcal{L}_{over}(\boldsymbol{x}) + \lambda_m \mathcal{A}_{margin}(\boldsymbol{x}) - \lambda_{area}\left(r_{1,2}^2 - r_{1,1}^2\right),$$

where the last term actively promotes geometric openness of the central channel. The weighting coefficients are chosen to balance the magnitudes of the loss terms and empirically yield stable convergence across diverse target spectra ( $\lambda_{over} = 10, \lambda_m = 1, \lambda_{area} = 0.03$ ). Geometric feasibility is incorporated directly into the hybrid framework to ensure physically realizable designs. The radial variables must satisfy strict monotonicity and thickness constraints:

$$r_{1,1} < r_{1,2} < r_{2,1} < r_{2,2}$$

$$r_{1,2} - r_{1,1} \geq \epsilon_1$$

$$r_{2,1} - r_{1,2} = \epsilon_2$$

$$r_{2,2} - r_{2,1} \geq \epsilon_3$$

where $\epsilon_2$ denotes the prescribed inter-region gap and $\epsilon_1$, $\epsilon_3$ impose minimum structural thickness constraints to prevent unrealistically thin features and excessive thermoviscous dissipation and fabrication failure. Rather than enforcing these conditions through soft penalties, a hard projection operator $\prod(\cdot)$ is embedded in the optimization loop, applied before and after each gradient update to ensure strict feasibility.

To bolster robustness, mitigate susceptibility to local minima, and enable efficient global exploration of the highly non-convex feasible design space, we adopt a population-based parallel optimization strategy. Let $N$ denote the number of initial distinct candidates, each parameterized as $x^{(n)} \in \mathbb{R}^6, n = 1, \dots, N$. The initial population is generated via grid-based sampling within prescribed geometric bounds to ensure diverse coverage. In practice, $N$ is set below 4000 to balance computational efficiency and population diversity.

Optimization proceeds through multiple parallel stages to accelerate convergence and progressively filter the candidate population. At stage $s$, the candidate set $X_s$ is updated in parallel using the Adam optimizer for $S_s$ iterations with a tailored learning rate $\eta_s$. Following optimization in each stage, candidates are first ranked in ascending order of their per-sample objective values $\mathcal{J}(x)$. The top $K_s$ distinct candidates are then selected via a greedy diversity-aware procedure. Specifically, candidates are traversed in order of increasing $\mathcal{J}(x)$, and each candidate $x^{(i)}$ is retained only if it is not geometrically similar to any previously selected, higher-ranked candidate (see Supplementary Information for details). The next-stage population is thus obtained as:

$$X_{s+1} = DiverseTopK_{K_s}(X_s)$$

where $DiverseTopK_{K_s}(\cdot)$ denotes the combined ranking and diversity-aware selection operator. Candidate similarity is exclusively evaluated over the radial parameter subset $(r_{1,1}, r_{1,2}, r_{2,1}, r_{2,2})$. Two candidates $x^{(i)}$ and $x^{(j)}$ are regarded as geometrically similar if

$$\left|x_d^{(i)} - x_d^{(j)}\right| > \delta_d, d \in \{1,2,3,4\},$$

where $\delta_d$ denotes the predefined tolerance for each geometric dimension. This stage-wise survivor selection concentrates computational resources on high-quality candidates while robustly preserving diversity to identify distinct design classes.

Upon completion of the optimization stages, a final hard-threshold filtering step is applied in the physical frequency domain to enforce strict band-limited attenuation. For each optimized candidate $x$, the predicted transmission spectrum $\hat{T}(f;x)$ is evaluated, and the maximum transmission within the original target band is defined as:

$$T_{\max}(x) = \max_{f \in [f_{\text{low}}^{\text{hard}}, f_{\text{high}}^{\text{hard}}]} \hat{T}(f;x)$$

Only candidates satisfying $T_{\max}(x) \leq \tau_{\text{hard}}$ are retained, where $\tau_{\text{hard}}$ is a prescribed strict transmission threshold. Further implementation details are provided in Supplementary Information.

Overall, the proposed PHP inverse prediction framework integrates population-based sampling, parallel surrogate optimization, hybrid-objective formulation, and stage-wise survivor selection to efficiently synthesize physically feasible UAS geometries that satisfy band-limited spectral constraints. A pseudocode summary of the complete procedure is provided in Algorithm 1. Compared with conventional finite element method (FEM) parameter sweeps that may require months of iterations, this framework drastically reduces the computational cost of structure synthesis. Once trained, the model enables inverse design of complex acoustic metamaterial structures within seconds per distinct design.

## 3. Result and Discussion

### 3.1. Interference Mechanism and Spectral Regimes

In this section, we elucidate the physical origin of the UAS performance by conceptualizing the bilayer structure as a subwavelength acoustic interferometer. Region 1 serves as the reference transmission continuum ($n_1 = 1$, $Z_1 = \rho_0 c_0$, where $\rho_0 = 1.22\ kg/m^3$ and $c_0 = 343\ m/s$), while Region 2 acts as a phase-delayed resonant pathway modeled via homogenization theory. The effective properties of Region 2 are dictated by the filling ratio $F$ (the volume fraction of the solid structure), which determines the relative bulk modulus $B_r = 1/(1-F)$ and the effective density $\rho_2 = \rho_0 n_2^2 B_r$. The waveguide radius is fixed at $R = 50$ mm and the structural thickness is set to $t_0 = 50$ mm. The interface between the two regions is separated by a small geometric gap such that $r_{2,1} = r_{1,2} + 0.5$ mm. This homogenization approach captures the complex acoustic response of the internal coiled geometry while preserving the analytical tractability required for high-throughput exploration.

Algorithm 1 Population-based Hybrid-objective Parallel Inverse Prediction

1: **Input:** Initial population $X_0$, surrogate model $g_\phi$, stage schedule $\{S_s, \eta_s, K_s\}_{s=1}^{S}$, diversity tolerances $\{\delta_d\}_{d=1}^{4}$, hard-threshold band $[f_{\mathrm{low}}^{\mathrm{hard}}, f_{\mathrm{low}}^{\mathrm{hard}}]$, hard threshold $\tau_{\mathrm{hard}}$

2: **Output:** Final filtered candidate set $X_{\mathrm{filt}}$

3: $X \leftarrow X_0$

4: for $s = 1,2,\ldots,S$ do

5: $X \leftarrow \mathrm{detach}(X)$

6: Set $\mathrm{requires_grad}(X) \leftarrow \mathrm{true}$

7: $\mathrm{optimizer} \leftarrow \mathrm{Adam}(X, \eta_s)$

8: **for** $\mathrm{iter} = 1,2,\ldots,S_s$ **do**

9: $\mathrm{ApplyFixed}(X, F)$ — overwrite fixed $t$, $\theta$

10: $X \leftarrow \Pi(X)$ — enforce hard constraints

11: $\hat{T} \leftarrow g_\phi(X)$

12: $\mathcal{J} \leftarrow Objective(\hat{T}, X)$

13: $\mathrm{optimizer.zero_grad}()$

14: $\mathrm{Backpropagate}(\mathcal{J})$

15: $\mathrm{ZeroGradFixed}(X, F)$

16: $\mathrm{optimizer.step}()$

17: $X \leftarrow \Pi(X)$ — re-project after update

18: $\mathrm{ApplyFixed}(X, F)$

19: **end for**

20: $\mathrm{score} \leftarrow \mathrm{PerSampleScore}(X, \hat{T})$

21: $X \leftarrow \mathrm{DiverseTopK}(X, score, K_s, \{\delta_d\}_{d=1}^{4})$ — diversity-aware survivor selection

22: **end for**

23: $\hat{T} \leftarrow g_\phi(X)$

24: $X_{\mathrm{filt}} \leftarrow \{x \in X \mid \max_{f \in [f_{\mathrm{low}}^{\mathrm{hard}}, f_{\mathrm{high}}^{\mathrm{hard}}]} \hat{T}(f; x) \le \tau_{\mathrm{hard}}\}$ — final hard-threshold filtering

25: **return** $X_{\mathrm{filt}}$

To generalize the geometric influence across different structural scales, we define the radial parameters using the dimensionless cross-sectional area coefficients:

$$k_{1,1} = r_{1,1}^2/R^2$$

$$k_{1,2} = (r_{1,2}^2 - r_{1,1}^2)/R^2$$

$$k_{2,2} = (r_{2,2}^2 - r_{2,1}^2)/R^2$$

Physically, $k_{1,1}$ represents the accessible internal area within the central channel, while $k_{1,2}$ and $k_{2,2}$ characterize the effective filling ratios of the continuum (Region 1) and the phase-delayed resonant pathway (Region 2), respectively. The ratio between these coefficients directly modulates the acoustic impedance matching at the interface, thereby governing the energy distribution and interference depth of the UAS.

To resolve the governing interference mechanisms, we performed systematic parameter sweeps across the descriptor space **(Figure 3a-e)**. The spectral response is interpreted through the normalized frequency $f' = n_2 t f / c_0$, which can be partitioned into four distinct interference regimes based on characteristic phase-accumulation boundaries: Regime 1 ($0 < f' < 0.5$), Regime 2 ($0.5 < f' < 1.0$), Regime 3 ($1.0 < f' < 1.5$), and Regime 4 ($1.5 < f' < 2.0$).

To quantify the interference efficiency within these regimes, we evaluate the system using two rigorous physical criteria: the velocity contrast and the phase-mismatch condition. The velocity contrast, defined as $|U_{1t} - U_{2t}|/\pi R^2$, measures the amplitude parity between the continuum and resonant pathways. Destructive interference reaches its theoretical limit only when the volume velocities of the two channels are perfectly balanced. Simultaneously, the phase-mismatch condition, $|(\varphi_{1t} - \varphi_{2t})/\pi - 1|$, monitors the departure from the ideal anti-phase requirement. Our analysis reveals that the transmission minima ($T \approx 0$) are strictly localized at the intersection where the velocity contrast vanishes and the phase mismatch approaches zero.

The broadband suppression observed across these regimes is inherently rooted in Fano-resonant interference. Region 1 provides a continuum background, while Region 2 introduces a spatially dispersive, discrete resonant pathway. Their destructive recombination at the outlet yields a characteristic asymmetric Fano profile, where the cross-section area ratio $k_{2,2}/k_{1,2}$ acts as a physical tuning knob for coupling strength. By balancing velocity amplitudes and synchronizing a $\pi$-phase mismatch, the system generates profound transmission dips. This mechanism allows the UAS to achieve steep spectral transitions and deep suppression within an ultra-thin profile, surpassing the limits of symmetric Lorentzian resonances.

Crucially, this physical mapping reveals a functional decoupling within the design space. The product $n_2 t$ acts as a global scaling factor that dictates the resonance's position. While variations in $t$ affect the phase accumulation across all regimes, $n_2$ primarily modulates the higher-order interference in Regimes 3 and 4. Collectively, these spectral parameters determine the operating frequency by shifting the attenuation bands without altering the underlying interference.

In contrast, the radial parameters $\{k_{1,1}, k_{1,2}, k_{2,2}\}$ function as lineshape modulators that govern the bandwidth and symmetry without shifting the normalized coordinates. Specifically, $k_{1,1}$ redistributes modal strength within the high-frequency Regime 4, while $k_{1,2}$ and $k_{2,2}$ serve as the primary drivers for broadening the effective attenuation regions across all regimes. This distinct separation of roles—where $n_2, t$ control the spectral location and radial geometry control the interference strength—provides the physical justification for our hierarchical two-stage surrogate architecture, enabling the model to resolve global spectral trends and localized resonant features independently.

Based on these behaviors, the optimized UAS designs can be physically classified into three categories according to their normalized transmission profiles, as shown in Figure 3f-h. UAS-1 exhibits a baseline resonant state in Regime 2 characterized by sharp attenuation but finite minimum transmission due to incomplete destructive interference. UAS-2 shows an optimized destructive interference in Regime 2, achieving near-perfect phase cancellation and near-zero transmission over a widened suppression band. UAS-3 presents an advanced multi-mode interference that coalesces multiple attenuation dips across Regimes 2 and 3, resulting in extended broadband suppression enabled by larger accumulated phase delay and enhanced modal coupling.

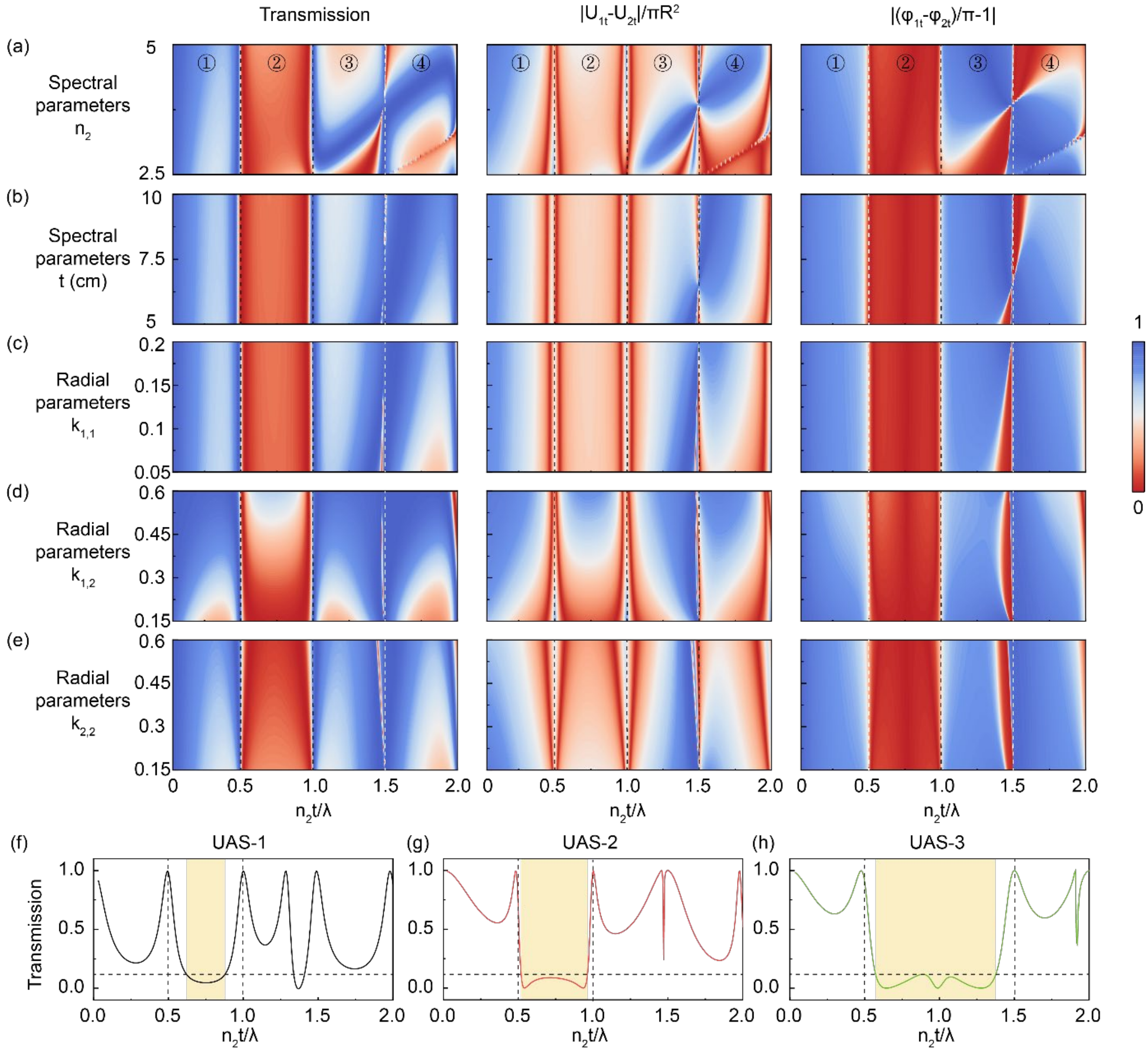

**Figure 3. Parameter-resolved physics and spectral classification of the bilayer interference system.** (a-e) Transmission colormaps (left), velocity contrast $|U_{1t} - U_{2t}|/\pi R^2$ (middle), and phase-mismatch $|(\varphi_{1t} - \varphi_{2t})/\pi - 1|$ (right) obtained by sweeping spectral parameters $(n_2, t)$ and radial parameters $(k_{1,1}, k_{1,2}, k_{2,2})$. The four spectral regimes are demarcated by phase-accumulation boundaries at intervals of $0.5n_2t/\lambda$. The maps reveal a fundamental functional decoupling: the product $n_2t$ serves as a global scaling factor for spectral positioning, while radial parameters act as lineshape modulators for interference bandwidth and depth. (f-h) Representative transmission spectra for three evolutionary UAS categories: UAS-1 (fundamental interference state in Regime 2), UAS-2 (optimized broadband suppression in Regime 2), and UAS-3 (multi-mode interference coalescing across Regimes 2 and 3). Yellow shaded areas denote the target silencing bands ($T < 0.1$).

### 3.2. Inverse Prediction Results on Spectral Parameters

To evaluate the design capability of the PHP framework, we performed inverse predictions to identify UAS geometries that synergistically optimize acoustic attenuation and ventilation capability. We introduce the effective angle $\theta_{\text{eff}} = \sin^{-1}(1/n_2)$ as a primary metric for aerodynamic performance. Physically, a larger $\theta_{\text{eff}}$ corresponds to reduced geometric tortuosity and lower airflow resistance, while the open-area ratio (determined by $k_{1,2}$) dictates the effective airflow cross-section. The inverse design task is thus formulated as a two-fold optimization: (i) maximizing the effective angle $\theta_{\text{eff}}$ for a specified thickness $t$, and (ii) maximizing the structural openness for a given $(t, \theta_{\text{eff}})$ configuration.

We first navigate the spectral design space by varying the thickness $t$ from 4 to 10 cm and the effective angle $\theta_{\text{eff}}$ from 10° to 30°, while maintaining representative radial constraints (e.g., $0 < r_{1,1} < 1\ cm$, $4 < r_{2,2} < 5$ cm). The target attenuation band is set to 1000-2000 Hz, with optimal solutions defined as designs achieving $T < 0.1$ over more than 75% of the bandwidth. For each point in the $(t, \theta_{\text{eff}})$ space, the PHP algorithm initiates a parallel search from 2,574 candidates, employing a stage-wise survivor selection strategy to filter and refine 100 high-performance broadband designs within minutes.

**Figure 4a** presents the analytical feasibility map derived from the $n_2 t/\lambda$ phase-accumulation criteria. The yellow domain ($0 < n_2 t/\lambda < 1.0$) primarily supports the fundamental interference modes (UAS-1 and UAS-2), while the green domain ($0 < n_2 t/\lambda < 1.5$) accommodates the higher-order multi-mode interactions characteristic of UAS-3. Regions where the broadband criterion cannot be physically satisfied are marked in grey, defining the fundamental limits of the design space.

The inverse prediction results (Figure 4b) show a remarkable alignment with these theoretical boundaries. The PHP framework successfully identifies both optimal (stars) and suboptimal (circles) candidates across the feasible region. Crucially, the red stars delineate the maximum achievable effective angle for each thickness. We observe that both the minimum and maximum angle limits required for broadband silencing increase monotonically with thickness $t$. This trend reveals a fundamental thickness-permeability trade-off: achieving broadband suppression in thinner profiles necessitates smaller effective angles (higher tortuosity), whereas thicker profiles allow for larger $\theta_{\text{eff}}$, thereby facilitating higher ventilation potential.

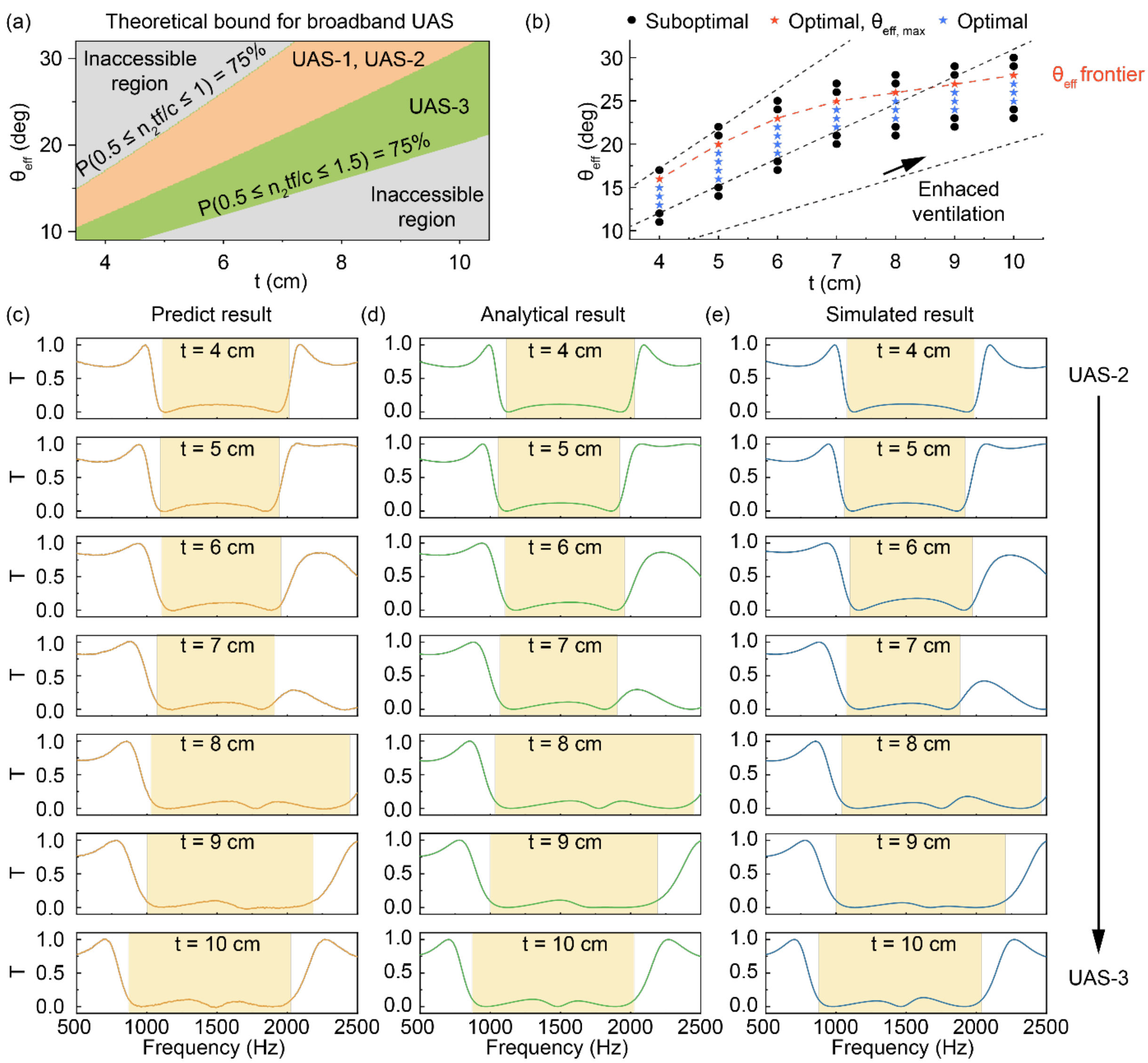

**Figure 4. Inverse design space and performance limits for spectral parameters.** (a) Analytical feasibility map of the $t$-$\theta_{\mathrm{eff}}$ space. Shaded regions indicate theoretical domains supporting broadband attenuation (> 75% coverage) based on phase-accumulation criteria: yellow for fundamental modes ($0 < n_2 t/\lambda < 1.0$, UAS-1/2) and green for higher-order interactions ($0 < n_2 t/\lambda < 1.5$, UAS-3). (b) Inverse prediction landscape. Stars and circles denote optimal and suboptimal broadband designs, respectively. The red stars delineate the optimal trade-off frontier representing the maximum achievable effective angle $\theta_{\mathrm{eff}}$ for a given thickness. (c-e) Comparison of transmission spectra for representative designs obtained via (c) surrogate model prediction, (d) analytical scattering theory, and (e) full-wave numerical simulations. The high-fidelity agreement across all three methods validates the framework's ability to resolve complex Fano features across distinct physical regimes.

The existence of a maximum $\theta_{\text{eff}}$ is fundamentally governed by the phase requirement. As $\theta_{\text{eff}}$ increases, the effective refractive index $n_2$ approaches unity, minimizing the accumulated phase delay $\Delta\varphi$ in Region 2. Below a critical $n_2$, the discrete pathway fails to provide the necessary $\pi$-phase mismatch relative to the continuum background within a subwavelength thickness $t$, rendering destructive interference physically unattainable. The red-star boundary thus represents the optimal trade-off between aerodynamic permeability and the acoustic phase-matching condition. Navigating this boundary requires the PHP framework to locate extremely narrow feasible regions in the radial parameter space, demonstrating its robustness in solving highly constrained physical problems.

The reliability of these inverse-designed solutions is validated through a three-way comparison between surrogate predictions (Figure 4c), analytical calculations (Figure 4d), and full-wave numerical simulations (Figure 4e). The exceptional agreement across all three methods confirms the framework's ability to accurately capture complex spectral envelopes and sharp Fano features. As the structure transitions from thinner to thicker profiles, we observe a clear physical evolution: thinner designs predominantly exhibit the near-perfect phase cancellation of UAS-2, while thicker designs, operating near the maximum $\theta_{\text{eff}}$ limit, transition into the multi-mode suppression regime of UAS-3. This transition, driven by the increased phase accumulation $n_2 t$, underscores the PHP framework's capability to traverse distinct physical regimes to satisfy stringent engineering constraints.

Notably, the PHP framework inherently addresses the "one-to-many" mapping problem, a notorious bottleneck in acoustic inverse design where distinct geometric configurations yield nearly identical transmission spectra. By maintaining a diverse population of candidates through its stage-wise survivor selection strategy, the algorithm identifies a broad spectrum of degenerate optima within the feasible design space, as evidenced by the dense solution clusters shown in **Figure 5a**. This capability transforms a traditional computational challenge into an engineering advantage: it allows the framework to provide a comprehensive library of candidate structures for a single target performance, empowering designers to select solutions that best satisfy secondary manufacturing constraints, such as minimum wall thickness or material volume fractions, without compromising acoustic efficiency.

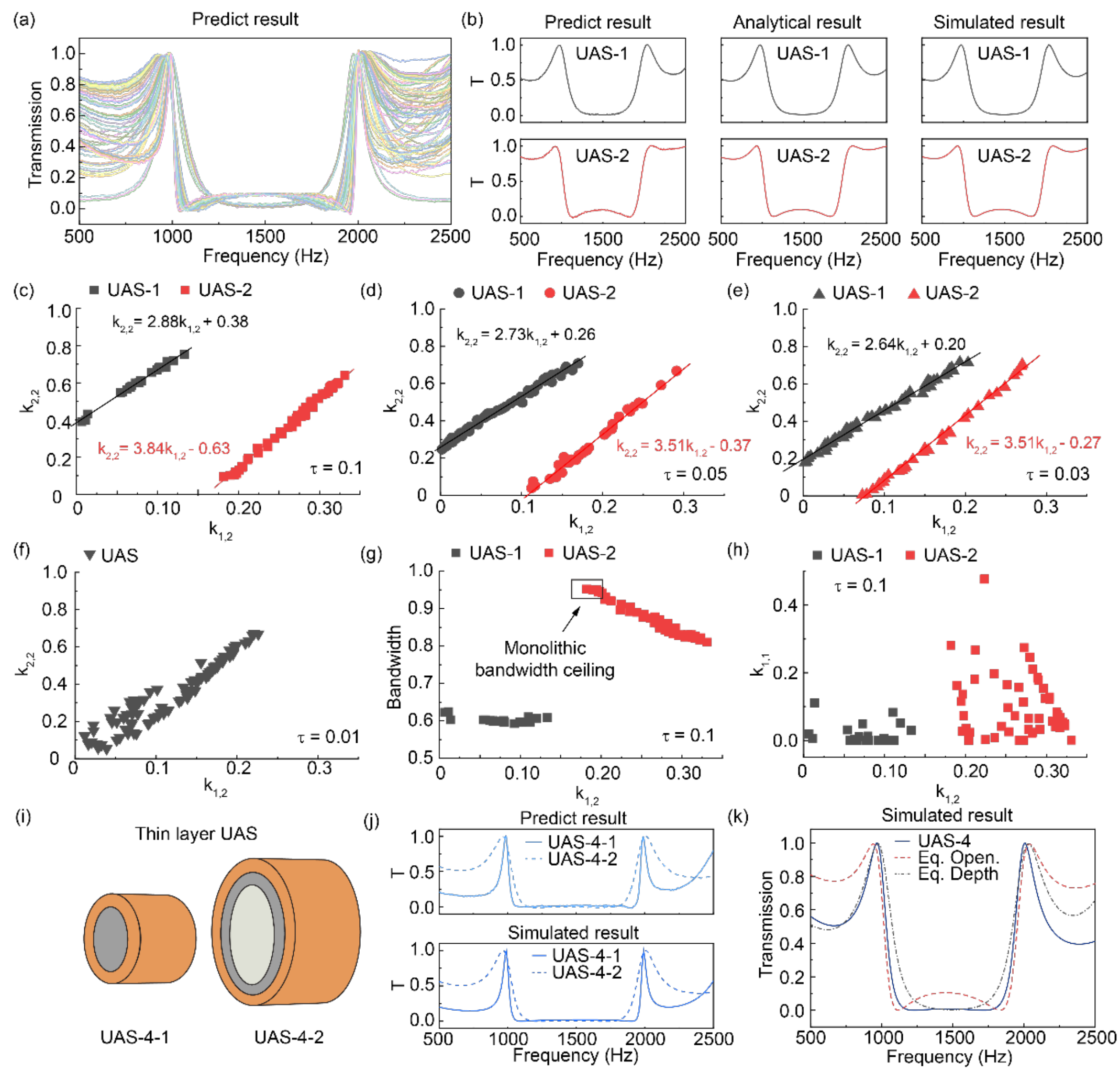

**Figure 5. Data-driven discovery of linear design rules within the radial parameter space and the evolution of the UAS-4 composite architecture.** (a, b) Collective transmission spectrum and representative benchmarks for UAS-1 and UAS-2 configurations. The high-fidelity overlap between surrogate predictions, analytical scattering theory, and full-wave simulations validates the framework's precision in capturing interference-driven dips. (c-f) Statistical mapping of the radial parameter space under varying attenuation thresholds $\tau$. The optimized designs cluster into two distinct linear trajectories ($k_{2,2}$ as function of $k_{1,2}$), revealing a hidden impedance-matching condition required for amplitude parity. As $\tau$ tightens from 0.1 to 0.01, the clusters converge toward a singular optimal physical state. (g, h) Performance trade-offs showing attenuation bandwidth and internal openness $k_{1,1}$ as functions of the continuum area $k_{1,2}$. The hollow square in (g) marks the bandwidth ceiling for monolithic designs. (i-k) Concept and performance of the composite UAS-4 architecture. By parallel integration of subunits (i, j), the UAS-4 assembly (k, solid blue curve) effectively circumvents the bandwidth-ventilation bottleneck, outperforming monolithic references with either equivalent openness (dashed red) or equivalent attenuation depth (dash-dotted black).

Complementing this versatility is the framework's exceptional computational efficiency. While generating an entire library of 100 high-performance designs across the design space requires only a few minutes, the inverse prediction for any single target design occurs within seconds. This performance represents a significant leap in design speed, several orders of magnitude faster than conventional iterative optimizers or genetic algorithms, thereby enabling the rapid, large-scale exploration of high-dimensional acoustic design spaces that were previously computationally prohibitive.

### 3.3. Inverse Prediction Results on Radial Parameters

To maximize structural openness, we navigate the radial design space $\{k_{1,1}, k_{1,2}, k_{2,2}\}$ for a fixed spectral configuration ($t = 4.7$ cm, $\theta_{\text{eff}} = 16°$). To prevent degenerate solutions with vanishing channel widths, we focus on the normalized frequency range $0.6 < f' < 0.9$. The PHP framework identifies a dense population of optimal candidates within this high-dimensional space, as shown in Figure 5a, where two dominant spectral signatures (UAS-1 and UAS-2) emerge. Representative spectra (Figure 5b) demonstrate exceptional agreement among surrogate predictions, analytical calculations, and full-wave simulations, confirming the framework's ability to resolve sharp, interference-driven attenuation features accurately.

A statistical analysis of the optimized solutions reveals that the geometric descriptors are not randomly distributed but instead cluster into two distinct linear regimes (Figure 5c). To quantify these characteristics, we extract the attenuation bandwidth (defined as the spectral fraction of 1000–2000 Hz where $T < 0.1$) and the cross-sectional area coefficients $\{k_{1,1}, k_{1,2}, k_{2,2}\}$ for each valid design. We observe that UAS-1 solutions primarily cluster within a low-openness range ($0 < k_{1,2} < 0.15$), whereas UAS-2 solutions occupy a broader high-openness domain ($0.15 < k_{1,2} < 0.35$).

Notably, the relationship between the coefficients $k_{1,2}$ and $k_{2,2}$ is not stochastic but follows two strict linear trajectories that define the operational boundaries of the system:

$$UAS - 1: k_{2,2} = 2.88k_{1,2} + 0.38$$

$$UAS - 2: k_{2,2} = 3.84k_{1,2} - 0.63$$

These dual-linear correlations serve as a geometric proxy for the optimal impedance matching condition required to sustain destructive Fano interference. Physically, the ratio of cross-sectional area coefficients $k_{2,2}/k_{1,2}$ dictates the volume velocity balance between the continuum and resonant pathways. Maintaining these specific proportions ensures that amplitude parity is preserved at the outlet as the structural scale varies, enabling deep attenuation across a wide range of geometric configurations. Remarkably, following these linear trajectories allows the system to broaden the ventilation ability while maintaining a near-constant peak attenuation, a feature that provides a practical "design ruler" for rapid geometric scaling without exhaustive re-optimization.

The robustness and sensitivity of these rules are further evaluated under varying attenuation thresholds, $\tau$ (Figure 5d–f). With relaxed thresholds ($\tau = 0.05$ and 0.03), the optimized solutions cluster into distinct, well-separated regimes. As the threshold is tightened, the UAS-1 (black) and UAS-2 (red) clusters migrate toward each other along their respective linear axes. Under an extremely stringent threshold ($\tau = 0.01$), these clusters eventually merge, indicating a convergence toward a singular, high-performance physical state where the distinction between interference regimes vanishes in favor of global optimality.

Furthermore, we identify an inherent bandwidth-openness trade-off. As shown in Figure 5g, the attenuation bandwidth is negatively correlated with $k_{1,2}$, suggesting that achieving extreme broadband performance necessitates a higher concentration of acoustic energy within narrower interference channels. Notably, UAS-2 designs demonstrate clear superiority in this trade-off: they simultaneously achieve broader bandwidths and significantly larger internal openness ($k_{1,1}$) compared to UAS-1 (Figure 5h). The maximum achievable bandwidth, identified by the hollow square in Figure 5g, represents the physical limit of monolithic architecture.

To bypass the bandwidth-ventilation bottleneck identified in monolithic designs, we introduce the UAS-4 composite architecture. The core strategy involves parallel integration of multiple smaller-area subunits, specifically UAS-4-1 (characterized by a smaller $r_{2,2}$) and UAS-4-2 (featuring a larger $r_{1,1}$), to achieve additive ventilation without compromising acoustic performance (Figure 5i). To mitigate potential performance degradation arising from mutual coupling between adjacent units, we employed a stringent inverse-design threshold of $\tau = 0.03$ during selection.

The individual performance of these subunits is shown in Figure 5j, where both UAS-4-1 and UAS-4-2 provide robust attenuation within the target band despite their restricted individual ventilation capacities. However, their synergistic combination in the UAS-4 structure achieves a remarkable unprecedented ventilation ratio while maintaining high-fidelity silencing. As demonstrated in Figure 5k, the combined UAS-4 assembly yields a broad silencing band (1095-1870 Hz) with a mean transmission of only 0.013.

The superiority of this composite approach is highlighted through a comparative analysis with monolithic reference designs. UAS-4 ($k_{1,2} = 0.30$, $k_{2,2} = 0.66$) significantly outperforms a monolithic reference with comparable openness ($k_{1,2} = 0.33$, $k_{2,2} = 0.64$; dashed curve in Figure 5k), which exhibits a higher average transmission of 0.05. Conversely, achieving UAS-4's attenuation depth in a monolithic design would require a threshold of $\tau = 0.01$, severely reducing ventilation ($k_{1,2} = 0.23$; dash-dot line in Figure 5k). By spatially distributing the interference sites, the UAS-4 architecture circumvents the causal constraints of single-cavity resonators, providing a scalable pathway for high-performance acoustic silencers with enhanced permeability.

## 4. Experiments and Performance Validation

To experimentally validate the proposed framework and dissect the underlying Fano interference physics, three representative ultra-open acoustic silencers, UAS-2 (monolithic), UAS-3 (multi-mode), and UAS-4 (parallel-composite), were fabricated via high-precision 3D printing **(Figure 6a-c)**. As indicated by the white arrows in the optical images, the structural implementation features a central air continuum (Region 1) surrounded by complex coiled-channel architectures (Region 2) designed to achieve the requisite effective refractive index $n_2$. Acoustic characterization was performed using a standardized four-microphone impedance tube, with the measured Transmission Loss (TL) spectra benchmarked against both full-wave simulations and PHP-framework predictions (Figure 6d–f).

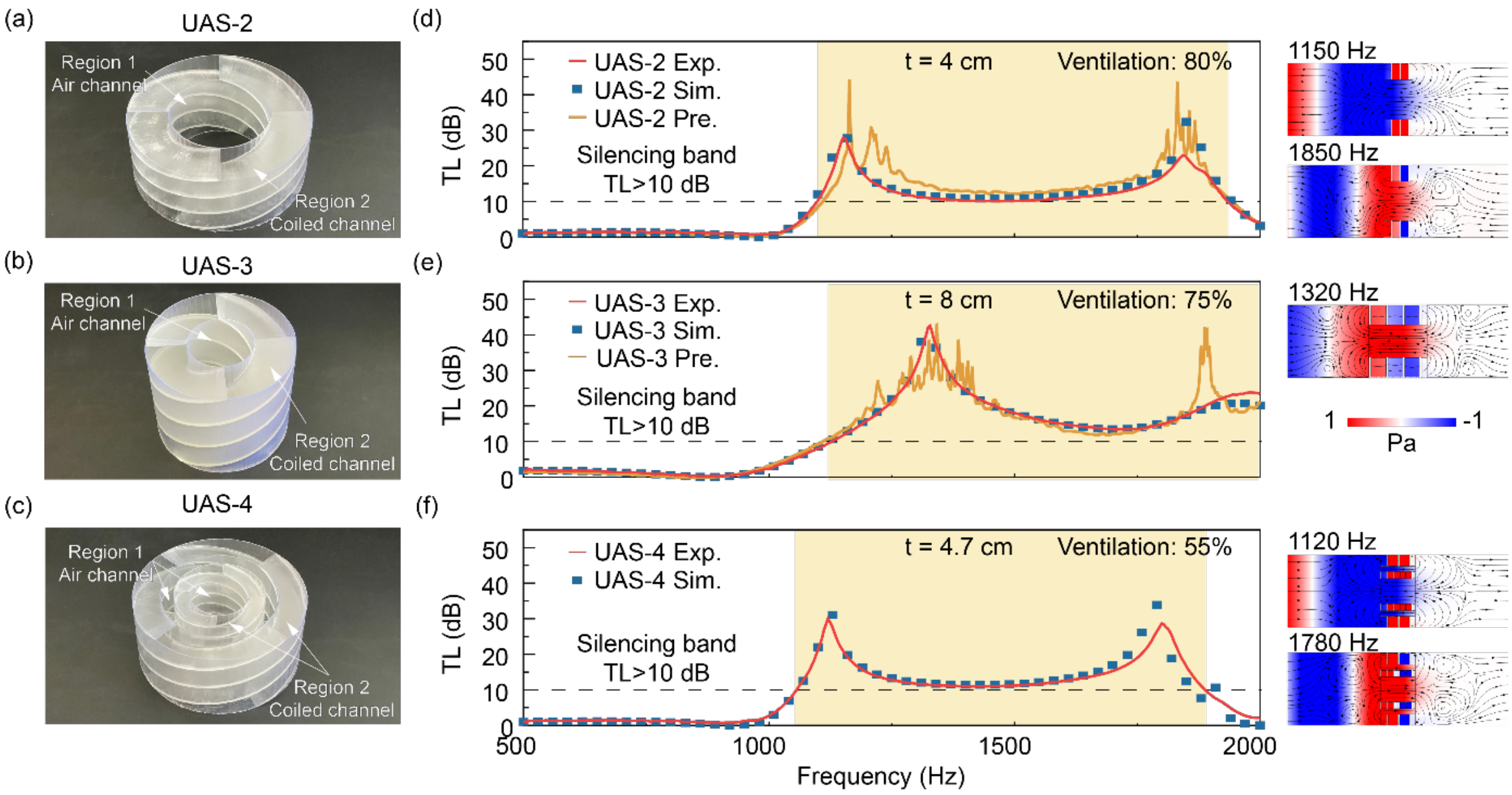

**Figure 6. Experimental validation and mechanistic discernment of the inverse-designed UAS prototypes.** (a-c) Optical images of the 3D-printed UAS-2, UAS-3, and UAS-4 structures. High-precision additive manufacturing enables the implementation of complex, coiled resonant pathways (Region 2) surrounding the central air continuum (Region 1), as indicated by the white arrows. (d-f) Quantified comparison of measured (solid red lines), full-wave simulated (blue symbols), and PHP-framework predicted (orange lines) Transmission Loss (TL) spectra for each design. The shaded yellow regions delineate the targeted broadband silencing bands (TL > 10 dB). All prototypes exhibit exceptional agreement between experiments and multi-way predictions, accurately capturing the Fano-resonant peaks and multi-mode interactions. Key performance metrics, including thickness $t$ and ventilation ratios (ranging from 55% to 80%), are highlighted in the upper right. Insets (Right): Simulated sound pressure field distributions and streamline patterns at the characteristic resonance frequencies for each design. The field maps provide direct visual evidence of destructive interference at the structural outlet, where phase mismatch and amplitude parity between the continuum (Region 1) and resonant (Region 2) pathways yield near-zero transmitted pressure, validating the underlying physical mechanism of the PHP-designed silencers.

The measured TL spectra (solid red lines) exhibit exceptional agreement with both numerical simulations (blue symbols) and surrogate predictions (orange lines), faithfully capturing the intricate, multi-resonance spectral envelopes across the entire 1000-2000 Hz target band. As highlighted by the yellow shaded regions, all three designs achieve robust broadband silencing ($TL > 10$ dB) exceeding 800 Hz. Quantitatively, UAS-2 ($t = 4$ cm) presents dual attenuation peaks at 1150 Hz and 1844 Hz (spanning 1099-1933 Hz), while UAS-3 ($t = 8$ cm) exhibits a sharp resonant dip at 1328 Hz (spanning 1128-2000 Hz). Notably, the composite UAS-4 ($t = 4.7$ cm) demonstrates pronounced peaks at 1120 Hz and 1800 Hz (spanning 1056-1888 Hz), maintaining deep attenuation even with its integrated parallel architecture. This multi-way consistency across monolithic, multi-mode, and composite architectures validates the high fidelity of the PHP framework in navigating the non-linear acoustic design landscape to identify optimal, physically realizable geometries.

Minor quantitative discrepancies in peak TL values are observed, though their origins differ. The deviation between experimental and predicted results is primarily amplified by the non-linear $TL = -10log_{10}(T)$ projection, where minuscule absolute errors in predicted transmission $T$ are magnified on the decibel scale. Meanwhile, the slight reduction in measured peak attenuation compared to lossless simulations is attributed to inherent thermo-viscous losses within the narrow-coiled channels, which mitigate the idealized interference dips. Crucially, the characteristic frequencies and silencing bandwidths remain nearly invariant across all three datasets, highlighting the inherent robustness of the Fano-resonant mechanism against both algorithmic approximations and boundary-layer dissipation.

To further substantiate the destructive interference mechanism, simulated sound pressure field distributions at characteristic resonance frequencies are presented to the right of each spectrum (Figure 6d-f). These field maps, accompanied by streamline patterns, provide direct visual evidence of Fano interference at the structural outlet. The near-zero transmitted pressure (white regions) results from the precise synchronization of amplitude parity and $\pi$-phase mismatch between the continuum (Region 1) and resonant (Region 2) pathways.

Table 1. Representative studies of ventilated acoustic silencers

| Number | Authors | Silencing band [Hz] | Ventilation ratio | Thickness [mm] |
|---|---|---|---|---|
| 1 | Dong et al. [20] | [628, 1400] | 34% | 53 |
| 2 | Yu et al. [58] | [700, 900] | 36% | 40 |
| 3 | Fu et al. [19] | [500, 3000] | 50% | 162 |
| 4 | Gao et al. [33] | [880, 1915] | 70% | 500 |
| 5 | Gao et al. [31] | [1040, 1464], [1525, 2365] | 55% | 42 |
| 6 | Yin et al. [25] | [1500, 1720] | 55% | 100 |
| 7 | Tang et al. [35] | [1145, 1815] | 32% | 100 |
| 8 | Sun et al. [34] | [900, 1418] | 20% | 100 |
| 9 | Dong et al. [59] | [610, 1120] | 20% | 50 |
| 10 | Zhu et al. [32] | [600, 1900] | 27% | 50 |
| 11 | Xu et al. [36] | [770, 1768] | 90% | 117.2 |
| 12 | Chen et al. [27] | [1025, 2000] | 39% | 30 |

Simultaneously, the aerodynamic performance was quantified by measuring the normalized airflow transmission ($v_w/v_{wo}$) at $v_{wo} \approx 4$ m/s. The designs achieved remarkable openness values of 80%, 75%, and 55% for UAS-2, UAS-3, and UAS-4, respectively. This dual-validation of broadband suppression (>800 Hz bandwidth) and unprecedented structural ventilation (up to 80% ventilation ratio) within a subwavelength thickness (0.1-0.2 $\lambda$) establishes a new benchmark for ventilated acoustic metamaterials. A quantitative benchmark comparison with previously reported silencers is summarized in Table 1. Conventional designs are typically bound by a severe broadband-openness trade-off, requiring either narrow channels (restricted ventilation) or increased thickness to achieve deep attenuation. In contrast, the proposed UAS designs simultaneously bypass these causal constraints, highlighting the advantage of the proposed PHP framework in realizing ultra-compact, highly-ventilated, and robustly silencing acoustic solutions.

## 5. Conclusion

In summary, we have established a physics-aware, machine-learning-assisted inverse design paradigm that transforms the design of complex, Fano-resonant acoustic silencers from exhaustive computation into a deterministic inverse-design process. By integrating effective-medium parameterization with a two-stage surrogate architecture, the framework effectively bridges the gap between physics-based modeling and high-dimensional data exploration. Specifically, the introduced two-stage forward prediction captures both broadband spectral envelopes and sharp resonant features through a coarse-to-fine strategy, while the population-based, hybrid-objective parallel (PHP) inverse strategy allows for the efficient navigation of the non-convex design landscape. This synergy enables the identification of geometries that simultaneously optimize broadband attenuation, structural openness, and geometric compactness.

Beyond its predictive capacity, the framework revealed hidden linear design rules that decouple spectral scaling from radial structural parameters. These rules serve as an optimal impedance-matching condition, providing a practical “design ruler” for scaling geometric openness without exhaustive re-optimization. Crucially, we introduced a parallel-composite architecture (UAS-4) to circumvent the bandwidth-openness trade-off ceiling inherent to monolithic resonators. By spatially distributing interference sites among smaller areas, the composite design achieves robust, deep attenuation while maintaining high ventilation capacity, demonstrating the framework's ability to discover paradigm-shifting architectural solutions.

Experimental validation of monolithic (UAS-2, UAS-3) and composite (UAS-4) prototypes confirmed the high fidelity of the PHP-derived designs. Near-perfect agreement between experiments, simulations, and predictions substantiated the Fano-resonant mechanism, with silencing bandwidths exceeding 830 Hz within 1000-2000 Hz and ventilation ratios up to 80% achieved within a subwavelength thickness (0.1-0.2 $\lambda$). These results establish a new performance benchmark for ventilated acoustic metamaterials and prove the inherent robustness of the proposed designs against algorithmic approximations and boundary-layer dissipation.

Overall, these results establish a general, physics-aware, machine-learning-assisted inverse design strategy for ventilated acoustic metamaterials, providing a systematic pathway to optimize the long-standing trade-off between acoustic attenuation and airflow transmission. By incorporating fundamental acoustic principles, specifically scattering-based phase requirements, into machine learning workflows, this platform unlocks a powerful toolkit for discovering multi-regime metamaterials that were previously computationally prohibitive. While validated here for broadband silencers, the workflow is inherently scalable; by integrating additional structural primitives, spatial configurations, and material parameters, it can be extended to a diverse range of acoustic wave-control systems. From industrial ventilation to aerospace applications, our work highlights the transformative potential of physics-informed, machine-learning-assisted inverse design to accelerate the deterministic discovery of next-generation wave-control devices.

## 6. Methods

All simulations were performed with finite element solver COMSOL Multiphysics 6.2 using the pressure acoustic module in the frequency domain. Ports are used to provide incident wave with certain amplitude and phase. The UASs were fabricated using stereolithography (SLA) 3D printing. A commercial 3D printer (Formlabs Form 4) with a printing resolution of 25 microns was employed. The transmission loss was measured using an impedance tube setup (100 mm B&K 4206-T large impedance tube). The sample was mounted in the sample holder, and four microphones were positioned in the upstream and downstream waveguides to characterize the acoustic response of the sample. To evaluate the ventilation performance of the UASs, additional experiments were conducted. In these measurements, an electric fan was placed at the inlet of the waveguide to generate airflow, and the wind speed at the outlet was recorded both with and without the presence of the UASs. The ventilation performance was quantified using the wind-velocity ratio, defined as the ratio of the airflow velocity measured with the sample to that measured without the sample.

**Acknowledgements**
The authors would like to thank Boston University Photonics Center and the Rajen Kilachand Fund for Integrated Life Science and Engineering for funding and technical support.

**Data Availability Statement**
Relevant data supporting the key findings of this study are available within the article and its Supplementary Information file. All raw data generated during the current study are available from the corresponding author upon request.

**Conflicts of Interest**
Provisional patent is filed based on the findings of this paper, detailed information is provided below. Provisional Patent Name: Materials and Methods for Ultra-Open Acoustic Silencers.

Inventors: Xin Zhang, Zhiwei Yang, Mengyu Li; U.S. Application No.: 64/047,672; BU Ref.: BU-2026-027; HBSR Docket No.: 6200.2034-000; File date: 04/23/2026.

**Author contributions statement**

Z.Y., M.L., and X.Z. conceived the study. Z.Y., A.C., and X.Z. conducted the numerical modeling and theoretical analysis. M.L., Z.Y., and X.Z. developed the code. Z.Y. and X.Z. fabricated the metamaterials. Z.Y., M.L., X.X., T.G.B., and X.Z. participated in building the experimental setup. Z.Y., M.L., and X.Z. designed and performed the experiments. All authors participated in discussing the results. Z.Y., M.L., and X.Z. wrote the manuscript. X.Z. provided funding acquisition, supervision, and resources.

**Supporting Information**

Supporting Information is available from the Online Library or from the author.